# Metallic Conduction at Organic Charge-Transfer Interfaces

Helena Alves, Anna S. Molinari, Hangxie Xie, and Alberto F. Morpurgo

*Kavli Institute of Nanoscience, Delft University of Technology, Lorentzweg 1, 2628 CJ Delft, the Netherlands*

**The electronic properties of interfaces between two different solids can differ strikingly from those of the constituent materials. For instance, metallic conductivity –and even superconductivity- have been recently discovered at interfaces formed by insulating transition metal oxides. Here we investigate interfaces between crystals of conjugated organic molecules, which are large gap undoped semiconductors, i.e. essentially insulators. We find that highly conducting interfaces can be realized with resistivity ranging from 1 to 30 kΩ/square, and that, for the best samples, the temperature dependence of the conductivity is metallic. The observed electrical conduction originates from a large transfer of charge between the two crystals that takes place at the interface, on a molecular scale. As the interface assembly process is simple and can be applied to crystals of virtually any conjugated molecule, the conducting interfaces described here represent the first examples of a new class of electronic systems.**

Many times in the past, the discovery of metallic electrical conductivity in molecular based materials has led to the development of new fields of research. Well known examples are iodine doping of polyacetilene [1], which started the field of polymer electronics, alkali intercalation of the fullerenes [2], with the subsequent discovery of superconductivity [3], and the development of charge transfer salts [4], in which a cornucopia of phenomena originating from electron-electron and electron-phonon interactions has been observed. All this work has been confined to bulk compounds and the realization of metallic conductivity at the surface of molecular-based materials has been so far much less successful (for very promising recent work in this direction see Ref. [5]). This is unfortunate, because the very rich electronic properties of surfaces and interfaces can enable the observation of new physical phenomena. In inorganic materials, for instance, this is illustrated by the discovery of metallicity [6] –and even superconductivity [7] and magnetism [8]- at interfaces between different insulating, non-magnetic transition metal oxides. Here, we show that a metallic system can also be created at the interface between crystals of two different conjugated organic molecules, which are both insulators in the bulk.

The molecules that we have selected for our study are Tetrathiofulvalene (TTF) and 7,7,8,8-tetracyanoquinodimethane (TCNQ), that are famous for the synthesis of the first metallic organic charge-transfer compound [9]. In such a compound, TTF and TCNQ are arranged in linear chains, forming a well-defined crystalline structure (Fig. 1). Electrons in the highest occupied molecular orbital (HOMO) of TTF are transferred to the lowest unoccupied molecular orbital (LUMO) of TCNQ, with virtually no hybridization between the molecular levels. As a result, the TTF and TCNQ chains behave as decoupled, one-dimensional electronic systems, and at room temperature the material is highly conducting. At low temperature, the compound becomes an insulator due to two Peierls transitions occurring independently on the TTF and TCNQ chains (at T=54 K for the TCNQ chain and T=38 K for the TTF chain), a clear manifestation of the underlying one-dimensionality of the material (for a review see [10, 11]).

Rather than focusing on bulk TTF-TCNQ, we have investigated electronic transport at the interface between TTF and TCNQ crystals. As shown in Fig. 2, TTF and TCNQ crystals are semiconductors, whose electronic properties are similar to those of many other organic materials (e.g., pentacene) commonly used in the field of plastic electronics, for instance for the fabrication of flexible field-effect transistors (FETs). The band-gap in these materials is approximately equal to the HOMO-LUMO gap of the corresponding molecule, which is larger than 2 eV [13]. As a consequence, the conductivity of pure TTF and TCNQ crystals is vanishingly small.

It is not a priori clear whether, similarly to the case of the bulk, charge transfer between the two molecules can indeed take place at a TTF-TCNQ interface. For this to happen, the energy of the charge-transfer state ($D^{\bullet+}A^{\bullet-}$, where D stands for donor and A for acceptor) has to be smaller than the energy of the system when the molecules remain neutral ($D^0A^0$). The energetic balance is determined by the difference in the energy of the TTF HOMO and the TCNQ LUMO –which is similar in the bulk and at the interface- and by the Coulomb energy stored in the charge transfer state, which is larger for a planar two-dimensional interface as compared to the interpenetrating linear chains present in the TTF-TCNQ bulk. It may therefore be that the additional cost in Coulomb energy prevents the occurrence of a large charge transfer. In addition, next to these fundamental aspects, technological problems may also prevent the assembly of a sufficiently ideal TTF-TCNQ interface. For instance, TTF-TCNQ interfaces may be unstable against mutual inter-diffusion of the two molecular species (a solid-state reaction between TTF and TCNQ could occur, resulting in bulk TTF-TCNQ[16]), or imperfection present at the crystal surfaces may trap all transferred charge.

The assembly of the TTF-TCNQ interfaces is done manually, in air, by first placing a TCNQ crystal onto a soft polydimethylsiloxane (PDMS) stamp, and by subsequently laminating a TTF crystal on top of TCNQ. The different steps involved are illustrated in Fig. 3. This lamination technique have been used extensively over the past years for the study of organic single-crystal field-effect transistors (FETs) [17], whose fabrication relies on the fact that thin organic single crystals of conjugated molecule adhere to the surface of virtually any insulating and metallic material. For instance, crystals of polyacenes (anthracene[18], tetracene[19], pentacene[20]), rubrene[21], metal phthalocyanines[22], thiophenes[23], and many other molecules, adhere easily onto insulators such as $SiO_2$, $Ta_2O_5$, $HfO_2$, $Al_2O_3$, $Si_3N_4$, PDMS [21,24], or onto metal films of Au, Pt, Ni, Cu, Co[25]. Organic single crystals also adhere spontaneously to molecular materials, such as self-assembled monolayers[20], PMMA[26], or other organic crystals. Recently, for instance, crystals of diphenylanthracene –which has a large HOMO-LUMO gap- have been laminated onto rubrene crystals and used as gate dielectric[27]. In all cases, adhesion occurs when the surface of the crystals is sufficiently flat, which is the case for most organic molecules[28] (for the substrate, the roughness has typically to be smaller than a few nanometers), and when crystals are sufficiently flexible to conform to the surface to which they are bonded. This is the case for ~ 1 micron thick crystals (similar to those used in this work), with the precise thickness being somewhat dependent on the molecular material. The mechanism responsible for the adhesion is not known in detail, but there is a wide consensus that the ubiquity of the phenomenon (i.e., the insensitivity to the materials used) points to electrostatic attraction (e.g., van der Waals forces), and that chemistry at the interface does not play a relevant role[17]. For the work presented here, it is particularly important that electrostatic bonding of organic crystals permits the reproducible fabrication of very high quality devices without introducing damage to the surface of the organic material[29], as demonstrated by a number of experiments. For instance, rubrene FETs fabricated by lamination have led to the discovery that, for high dielectric constant insulators, charge carriers in rubrene interact strongly with the self-polarization cloud in the insulator, resulting in the formation of interfacial Frohlich polarons[21]. Crystal lamination has also enabled reproducible contact resistance in organic transistors with Nickel electrodes, with record-low values (100 $\Omega$cm) [30]. However, the operation of single-crystal devices studied so far did not rely on the occurrence of charge transfer between two different materials, and it remains to be seen whether in-air crystal lamination can also be used for the assembly of the TTF-TCNQ interfaces that we are interested in.

We have investigated more than 50 TTF-TCNQ interfaces by performing transport measurements in air, in vacuum, and as a function of temperature. Whereas the resistance of individual TTF or TCNQ crystals is (much) larger than 1 GΩ at room temperature, for all TTF-TCNQ interfaces the value of resistance measured in a two-terminal configuration was typically lower than 100 kΩ. This value includes the contact resistance, which is in general large (see Fig. 4 a and b), and to measure the contribution of the TTF-TCNQ interface it is essential to employ a four terminal configuration. Owing to the small size of the samples (typically ∼ 500 μm linear dimension), we succeeded in attaching four contacts only to a fraction of the devices (approximately 15). From these devices we found that the resistance-per-square of the TTF-TCNQ interface ranges from 1kΩ/square to approximately 30 kΩ/square. The spread appears to be mainly related to the quality of the different devices, but, in part, it may also originate from anisotropy in the conduction. In fact, the lamination process used for the interface assembly does not give good control of the relative orientation of the two crystals, so that in different devices transport occurs along different crystalline directions (the mobility of charge carriers in organic crystals is known to be slightly anisotropic, typically a factor of 3-4, for materials with a molecular packing similar to that of TTF and TCNQ[24,31]). Note also that TTF-TCNQ interfaces were assembled using two different crystalline phases of TTF, which have different lattice constants (the α-phase and the β-phase; see Methods). High conductivity was observed in both cases, suggesting that its occurrence does not crucially depend on the commensurability of the TCNQ and TTF lattices. Finally, the conductivity values are essentially identical irrespective of whether measurements are performed in air or in vacuum (see Fig. 4c) and the interfaces were stable over long periods of time. This is illustrated by Fig. 4d, which shows that the conductivity does not change significantly over a period of months.

The measurements of the temperature dependence of the conductivity are of particular interest. For a two-dimensional conductor whose square resistance varies from being much smaller than, to comparable or larger than the resistance quantum ($h/e^2$ ∼ 26 kΩ), a cross-over in the observed temperature dependence should be expected [32]. Specifically, when the square resistance is high, electronic states tend to be localized and the interface becomes more resistive as the temperature is lowered. On the contrary, when the square resistance is much smaller than the quantum resistance, the resistance should decrease with lowering temperature in a broad temperature range. Indeed this is what is observed experimentally, as illustrated in Fig. 5c by measurements performed on three different samples: the temperature dependence of the resistance of an interface with R ~ 1 kΩ/square decreases with lowering temperature, whereas an interface with R ~ 6 kΩ/square at room temperature becomes more resistive upon cooling. In total we performed temperature-dependent measurements on five interfaces with resistance smaller than 10 kΩ/square, all consistent with the data shown in Fig. 5c and with two of them exhibiting a metallic temperature dependence (two of these five samples could only be measured in a smaller temperature range, because the crystals cracked during cooling). We also measured other devices with a larger square resistance, which exhibited insulating behavior, as expected.

Another key observation enabled by the temperature-dependent measurements is that no sharp resistance increase is seen throughout the entire temperature range. In particular, no special feature in the data is present at T=54 K, where in bulk TTF-TCNQ the resistivity increases by a factor of 20 in a few Kelvin range[10, 11] owing to the occurrence of the first Peierls transition. Our data shows unambiguously that, contrary to the bulk, such a transition does not occur at TTF-TCNQ interfaces. The absence of a Peierls transition allows us to conclude directly that the electronic properties of TTF-TCNQ interfaces are intrinsically different from those of bulk TTF-TCNQ. This excludes that the high conductivity measured at TTF-TCNQ interfaces originates from the formation of a bulk-like layer (possibly formed as a consequence of inter-diffusion of the two molecules), in which case the effect of a Peierls transition should also be visible. This finding is consistent with the observed stability of our samples, since material inter-diffusion would lead, after increasingly long periods of time, to an increasingly thicker conducting layer near the interface and, correspondingly, to a decrease in the measured resistance (which is not what we observe; see Fig. 4d). We conclude that at TTF-TCNQ interfaces a metallic conductor is formed due to charge transfer between the two molecules.

Having established the occurrence of metallic conductivity at TTF-TCNQ interfaces, we take a first step in the analysis of the interfacial electronic properties. An important aspect of bulk TTF-TCNQ is the negligible hybridization between the molecular orbitals of the TTF and of the TCNQ chains. Since the molecular planes are very weakly coupled already within individual crystals of TTF and TCNQ, it is likely that also at the TTF-TCNQ interface the coupling between the adjacent TTF and TCNQ planes is weak, i.e. the hybridization between orbitals of the two molecules is likely to be very small in this case as well. It follows that the interfacial conductivity that we observe is due to two decoupled conducting layers with equal and opposite amount of charge: one in which conduction is due to holes in the HOMO band of TTF, and the other in which current is carried by electrons in the conduction band of TCNQ. With the surface density of TTF and TCNQ molecules being approximately $5 \cdot 10^{14}$ cm$^{-2}$, and assuming the amount of charge transferred from TTF to TCNQ at the interface to be approximately the same as in the bulk -0.59 electron charges per molecule [10, 11]-, we obtain a square resistance of ~ 1 k$\Omega$ for mobility values of a few cm$^2$/Vs at room temperature (entirely realistic in our crystals). This value compares unexpectedly well with the square resistance that we measure in our best samples (the high resistance values measured in other samples can have many different origins, such as a non uniform adhesion of the crystals or the presence of defects reducing the mobility). The absence of a Peierls transition at the interface is also easy to understand. In fact, in crystals of conjugated molecules [13] whose structure is similar to that of TTF and of TCNQ, only a small in-plane anisotropy of the carrier is observed experimentally[24,31]. This implies that the electron and holes at the surface of the TCNQ and TTF crystals are not confined to move along one-dimensional chains, as it happens in bulk TTF-TCNQ. Finally, with the surface charge density being larger than $10^{14}$ carriers/cm$^2$, the Thomas-Fermi screening length is comparable or smaller than the size of a molecule. As a consequence, electron and holes at TTF-TCNQ interface do not penetrate significantly in the bulk of the organic crystals, and the conducting layer is only a few molecules thick.

Given the similarity of TTF and TCNQ to other conjugated molecules used in plastic electronics, it is interesting to compare our results to existing work on other related systems, in which electronic transport also takes place at the surface of organic materials. In organic FETs, for instance, the lowest resistivity measured to date is of the order of 100 k$\Omega$/square –between one and two orders of magnitude larger than the resistivity values reported here. These values are attained at surface carrier density between 1 and 5 $10^{13}$ carriers/cm$^2$, at mobility values around few cm$^2$/Vs [33]. Since in our TTF and TCNQ crystals the mobility is of comparable magnitude, the much lower resistivity of TTF-TCNQ interfaces originates from a much higher density of charge carriers. Indeed, our estimate of ~5 $10^{14}$ carriers/cm$^2$ (see above) is between one and two order of magnitude larger than the largest carrier densities attained in FETs. Similar considerations hold true for structures in which charge transfer from other molecules was previously employed to control the surface density of charge carriers. One example is provided by organic FETs in which self-assembled monolayers chemically bonded to the gate dielectric are used to tune the threshold voltage (see, e.g., Ref. [34]). The amount of charge transferred by this technique, however, is low, typically few times $10^{12}$ carriers/cm$^2$. Much higher carrier densities have been recently achieved by direct chemical bonding of a self-assembled monolayer onto the surface of rubrene crystals [5]. Still, also in this case, both the carrier density and the conductivity were found to be between one and two orders of magnitude lower than in TTF-TCNQ charge transfer interfaces. Overall, therefore, the comparison between the conductivity at the surface of organic materials measured in different configurations and that of TTF-TCNQ interfaces, quantitatively supports our conclusion that in this latter system the carrier density is much higher.

More in general, the electron-hole metallic conductor that is formed at the TTF-TCNQ interface is a new type of electronic system. We believe that in charge transfer interfaces of this kind –based on TTF and TCNQ or on other similar pairs of conjugated organic molecules-, new physical phenomena may appear due to electron correlations, which in organic materials are known to be strong due to the narrow electronic bandwidth. For instance, in quasi two-dimensional organic charge transfer salts based on the BEDT-TTF molecule, superconductivity occurs at temperature as high as 10 K at ambient pressure[11]. Since BEDT-TTF based materials are formed by stacked molecular planes that are electronically decoupled, it seems definitely possible that superconductivity will also occur at charge transfer interfaces assembled in the way described here. Another interesting electronic state that could be formed at low temperature is an excitonic insulator[35], which is predicted to be realized in systems consisting of an electron and one hole layer closely facing each other with a negligible probability of inter-layer tunneling (precisely as in our TTF-TCNQ interfaces). Next to these specific examples, the virtually infinite classes of organic molecules that can be used to assemble molecular interfaces suggests that organic charge transfer interfaces will offer an unprecedented versatility for the realization of two-dimensional conductors with tunable electronic properties (e.g., molecules with spin could be employed to induce and control magnetic phenomena). In the field of organic charge transfer salts, a myriad of molecules have been synthesized over the last two decades in order to control the properties of new molecular materials, often to find that the co-crystallization of these molecules in a single compound is extremely challenging or prohibitively difficult. Working with interfaces – rather than with bulk materials- has the potential to solve this problem.

**Methods**

TCNQ crystals were grown by a vapor phase transport method commonly used for organic materials [36], with Argon as a carrier gas, at ambient pressure. TTF crystals were grown both from vapor phase and from solution. For this material, vapor phase growth is possible under mild vacuum (~ 1 mbar), which is needed to lower the temperature required to sublime TTF molecules at a sufficient rate, without inducing chemical decomposition. β-phase single TTF crystals are obtained in this way. TTF crystals were also grown from solution by dispersing 150 mg of material in 50 ml of *n*-hexane, and letting the solvent evaporate slowly over the course of two weeks. In this way α-TTF crystal are obtained. Both α- and β-TTF-TCNQ interfaces are highly conducting, but from the statistical analysis of our results, it appears that α-phase TTF-TCNQ interfaces exhibit on average slightly higher conductivity than β-phase TTF-TCNQ interfaces. It is currently unclear if this difference is due to an intrinsic property of the interface or from other factors such as, for instance, differences in the adhesion of the crystals.

As mentioned in the text, the interfaces are mounted on a PDMS support. The use of PDMS –as opposed to a rigid substrate such as glass or Silicon- is essential to perform low temperature measurements below 200 K. In fact, when mounted on rigid substrates, the crystals crack upon cooling, due to the different thermal expansion coefficients of the molecular materials and of the substrate. Electrical contact to the interfacial conducting layer is achieved by using a conducting carbon paste consisting of a water based suspension of carbon nano-particles, which is manually deposited (under an optical microscope) at the edges of the two crystals, to wet the interface. With practice, this technique enables the fabrication of contact with size smaller than 50 microns. A 25 micron gold wire is subsequently attached to the carbon paste protruding away from the interface using a solvent-free silver epoxy, and connected to the (standard) chip carrier onto which the devices are mounted.

Electrical measurements were done both in air and in the vacuum chamber of a flow cryostat with essentially identical results, using either a HP 4156A or an Agilent Technologies E5270B parameter analyzer. Measurements as a function of temperature require very slow cooling, since large and abrupt changes in temperature can result in the formation of cracks in the crystals, even when the crystals are mounted on PDMS supports. In practice, we often took an entire day to measure I-V curves of a device down to the lowest temperature in the experiments, starting from room temperature. In the cryostat that we have used to ensure sufficiently low cooling rates, non ideal radiation shielding from the cryostat walls prevents the sample temperature to be lowered below 40 K. Future experiments will aim at reaching lower temperatures and at applying a large magnetic field.

**Acknowledgments**

One of us (AFM) gratefully acknowledges a useful conversation with Dirk van der Marel. H. Alves acknowledges FCT for financial support under contract nr.

SFRH/BPD/34333/2006. Financial support from NanoNed and NWO is also acknowledged.

**Figures**

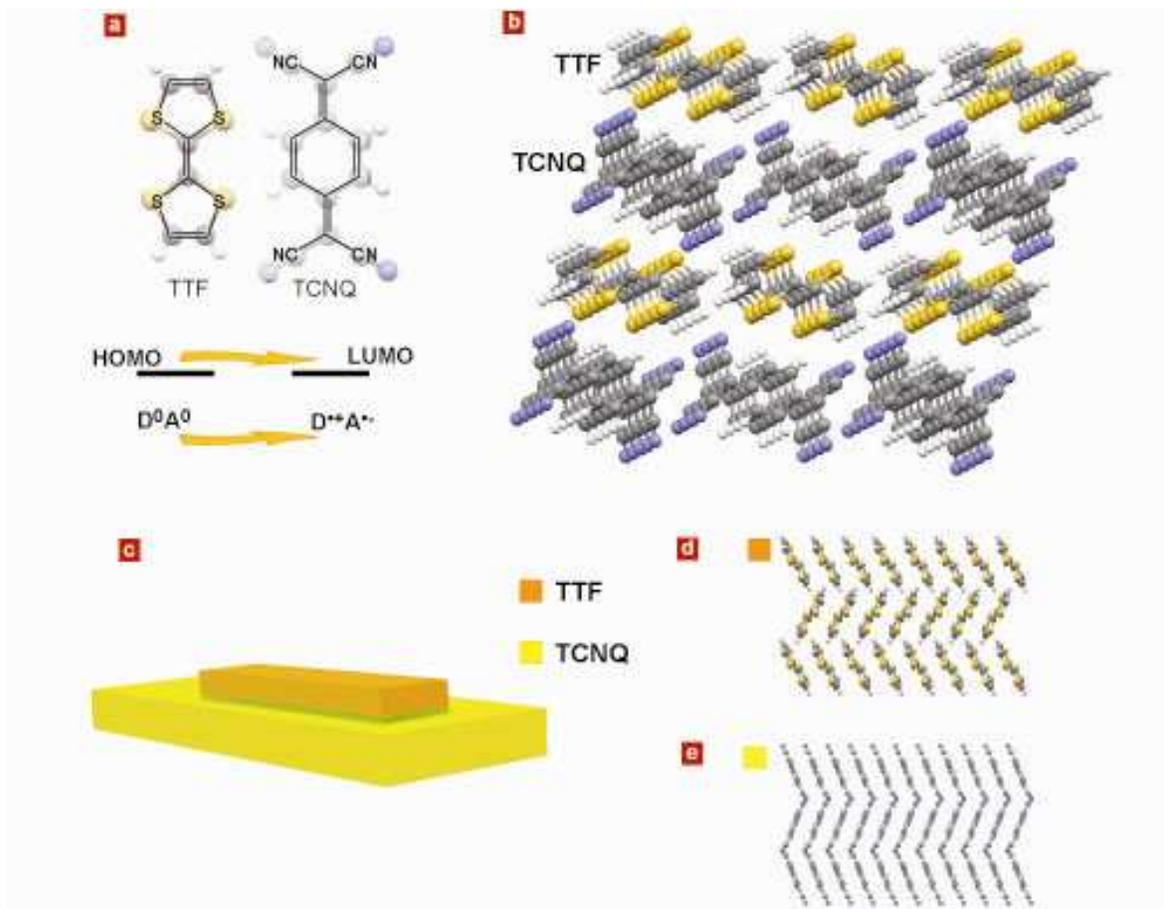

Figure 1 **Charge transfer in the TTF-TCNQ system.** The TTF and TCNQ molecules (panel a) are well-known since their use in the synthesis of the first metallic charge transfer compound, TTF-TCNQ, the structure of which is shown in panel b) [12] illustrating the quasi one-dimensional, chain-like arrangement of the TTF and TCNQ molecules. In TTF-TCNQ crystals, electrons from the HOMO of the TTF molecules are transferred into the LUMO of the TCNQ molecules, leading to a stable charge transfer state (as indicated in the diagram in panel a). Our work investigates whether a similar charge transfer occurs at the interface between a TTF and TCNQ crystal (blue shaded region in panel c), leading to the occurrence of a metallic state. Panel d) and e) show a top view of the structure of the molecular planes (note the characteristic herringbone configuration) where conduction takes place in (α-phase) TTF[14] and TCNQ[15], respectively. In our samples these molecular planes are parallel to (and form) the TTF-TCNQ interface. All structures shown are courtesy of the Cambridge Structural Database.

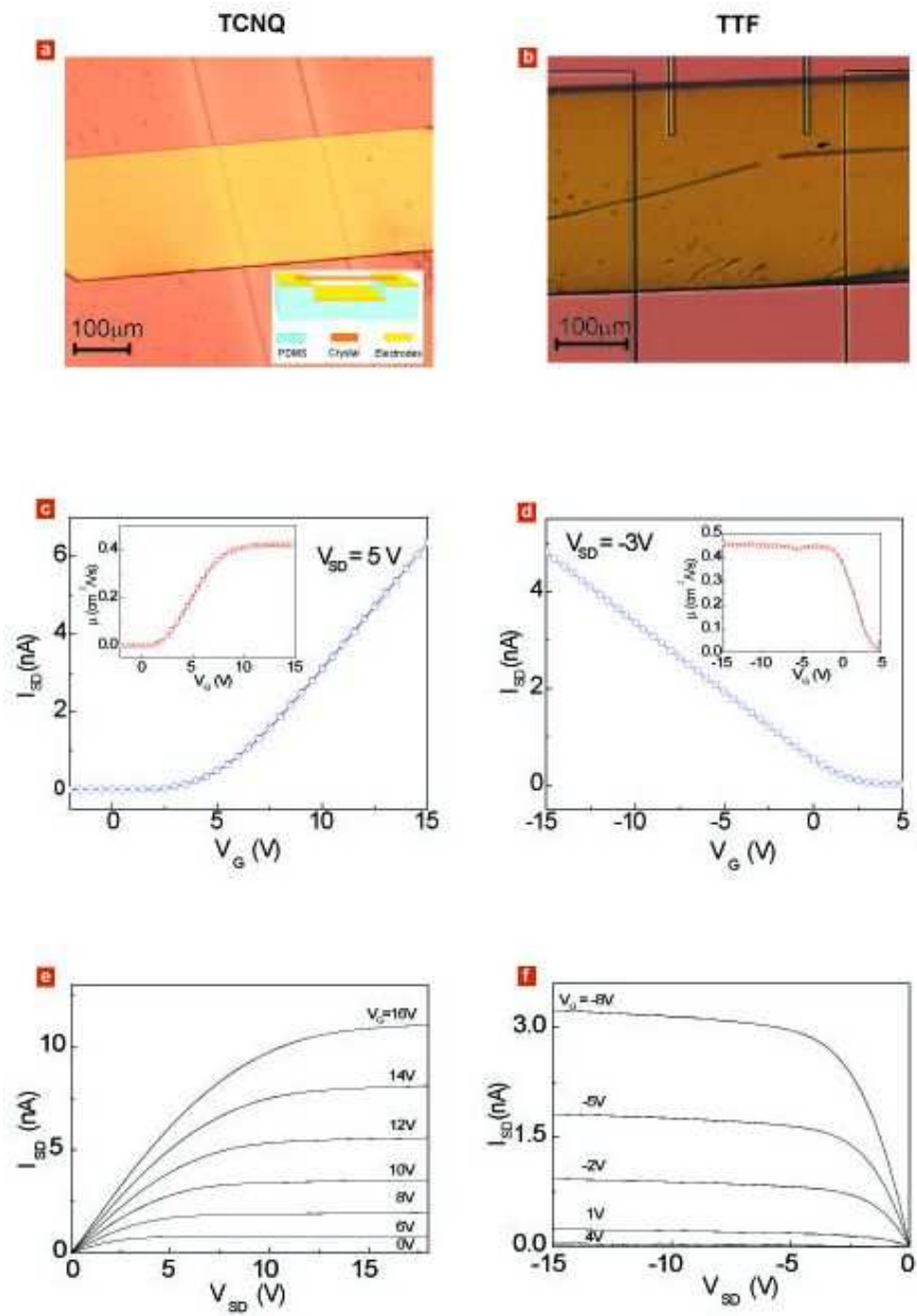

Figure 2 **Characterization of TCNQ and TTF single crystals.** The electrical characterization of the TCNQ and TTF single crystals used in assembly of TTF-TCNQ interfaces relies on the use of field-effect transistors (FETs). The devices are fabricated by laminating a single crystal onto a so-called PDMS air-gap stamp, as shown in the inset

of panel a) (for technical details see Ref. [37]). An optical microscope image of a single-crystal TCNQ FET is shown in panel a). Panel c) is a plot of the source drain current $I_{SD}$ measured on one of our TCNQ devices, as a function of gate voltage and for a fixed source drain bias $V_{SD}$=5 V. As expected for TCNQ –which is an electron acceptor- the current increases for positive gate voltages, i.e., when electrons are accumulated in the channel. In the off state (negative gate voltages) the device resistance is much larger than 1 GΩ, as expected since TCNQ crystals are undoped, large gap semiconductors. The inset shows the electron mobility of the same device obtained by numerically deriving the I(V) curve. Electron mobility values as high as 2 cm$^2$/Vs have been observed in vapor phase grown TCNQ single crystals (see, e.g., [37]); the value here is lower, probably because of a non negligible contact resistance at the source and drain electrodes is present in our two-terminal configuration. Panel e) illustrates the transfer curves of the device measured at different (positive) gate voltages, showing ideal transistor characteristics. Panels b),d), and f) provide analogous information for TTF single-crystals. Note that for TTF FETs, contrary to the case of TCNQ devices, conduction is observed for negative gate voltages since charge carriers are positively charged holes. Also for TTF crystals, the background resistance in the off state is much larger than 1 GΩ. All measurements of FET characteristics were performed at room temperature and in the vacuum (~10$^{-6}$ mbar).

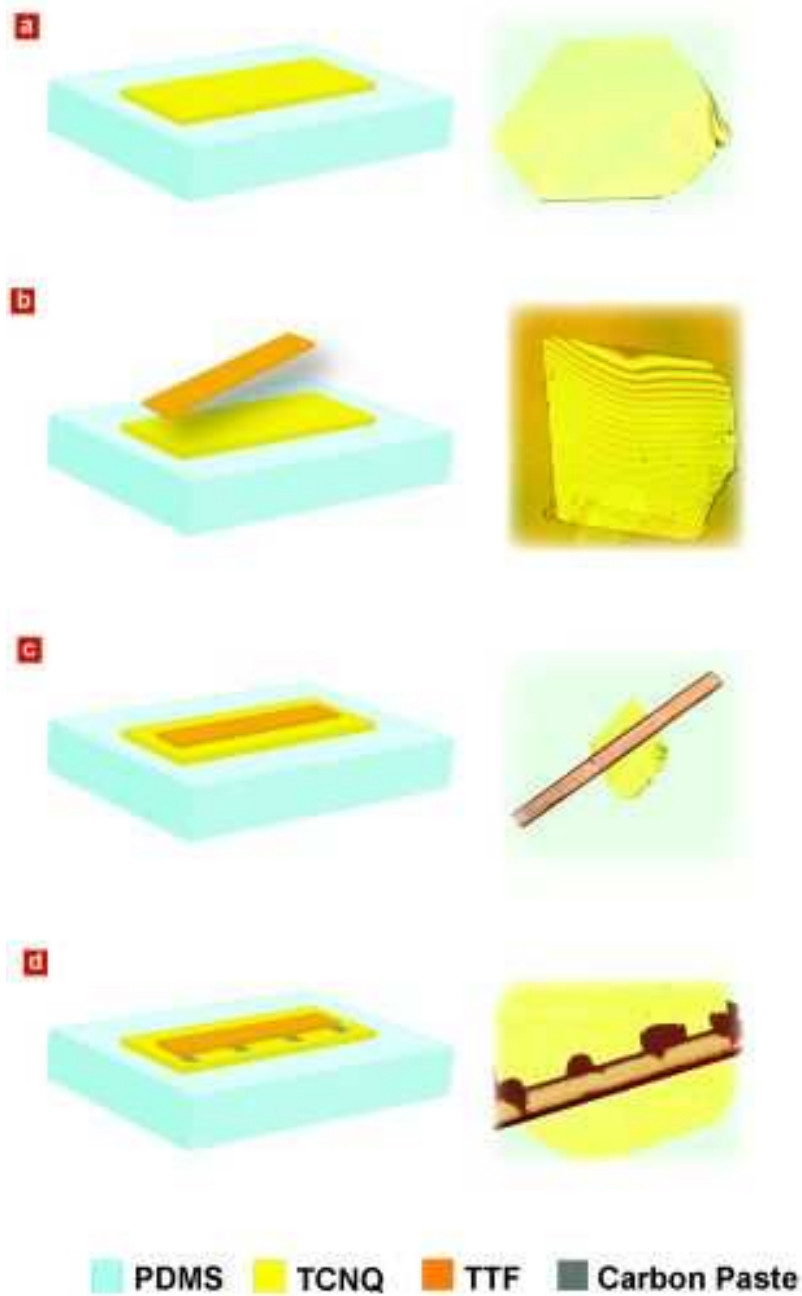

Figure 3. **Assembly of TTF-TCNQ charge transfer interfaces.** This figure illustrates the different steps followed to assemble TTF-TCNQ interfaces: schematics drawing of the individual steps are shown in the left column and the corresponding optical microscope images in the right column. In the first step (panel a), a TCNQ single crystal is placed onto a PDMS substrate. A thin TTF crystal positioned on top spontaneously adheres to the TCNQ crystal (panel b). The interference fringes seen in the optical

microscope image are characteristically observed during this adhesion process, and they disappear when the adhesion is complete. Panel c) shows the result of the crystal bonding process. Electrical contacts to the interface are fabricated using a carbon paste consisting of a water-based suspension of amorphous carbon nanoparticles, which are deposited manually at the edge of the interface (panel d). Gold wires for electrical measurements are attached to the carbon paste using a solvent-free silver epoxy.

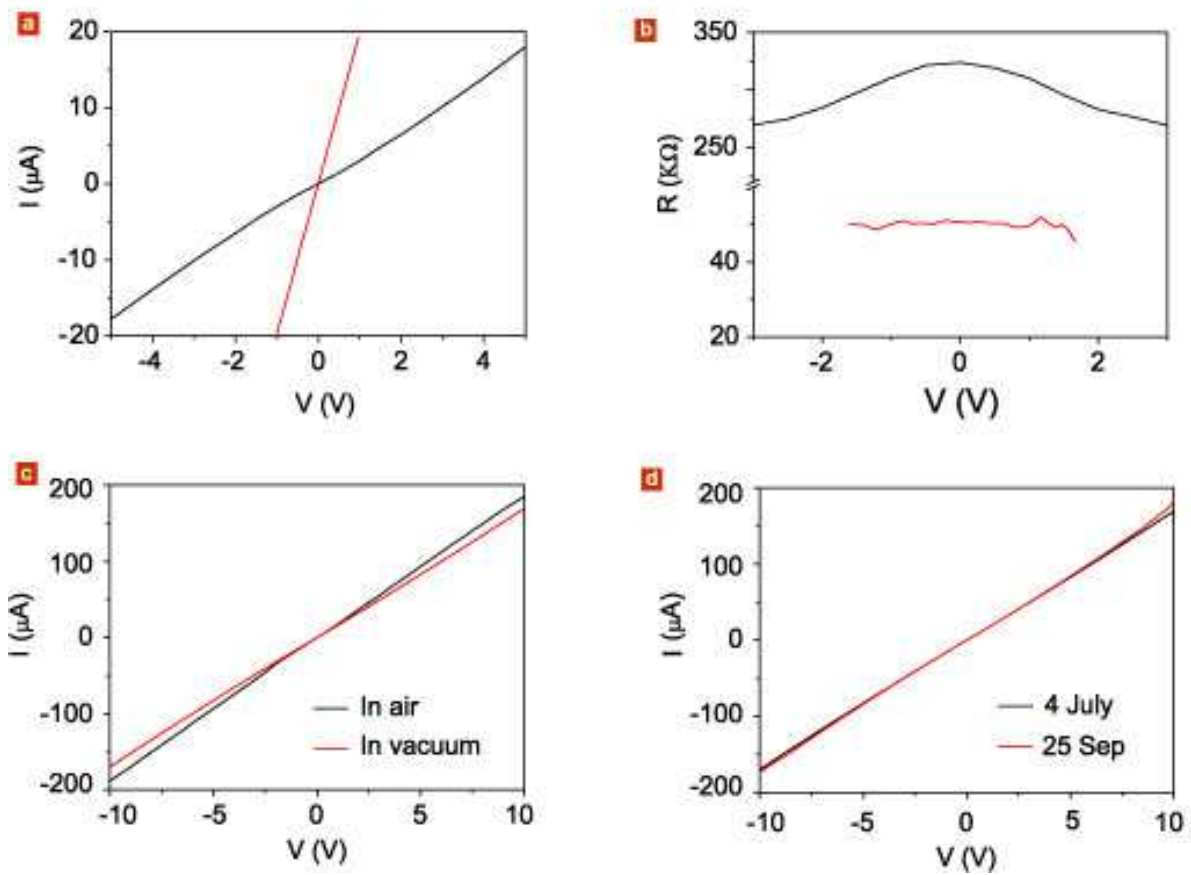

Figure 4. **Room temperature electrical characteristics of TTF-TCNQ interfaces.** All interfaces that we have measured –more than 50- exhibit a large electrical conductance, typically six orders of magnitude (or more) larger than the conductance of the individual TTF and TCNQ crystals. Panel a) shows a comparison between the I-V curves measured on a same device, in a two-terminal configuration (black curve) and in a four-terminal configuration (red curve). As it can be seen from the comparison, the contact resistance is in general large, and it typically results a characteristic non-linearlity a low bias. The non-linearity is more clearly visible in the differential resistance (panel b) and is completely absent in the four-terminal measurements. Panel c) shows that a high conductance is observed irrespective of whether measurements are performed in air or in vacuum. The TTF-TCNQ interfaces assembled by lamination are very stable as shown by the measurements in panel d), where no difference is observed in the measurements performed almost three months after fabrication.

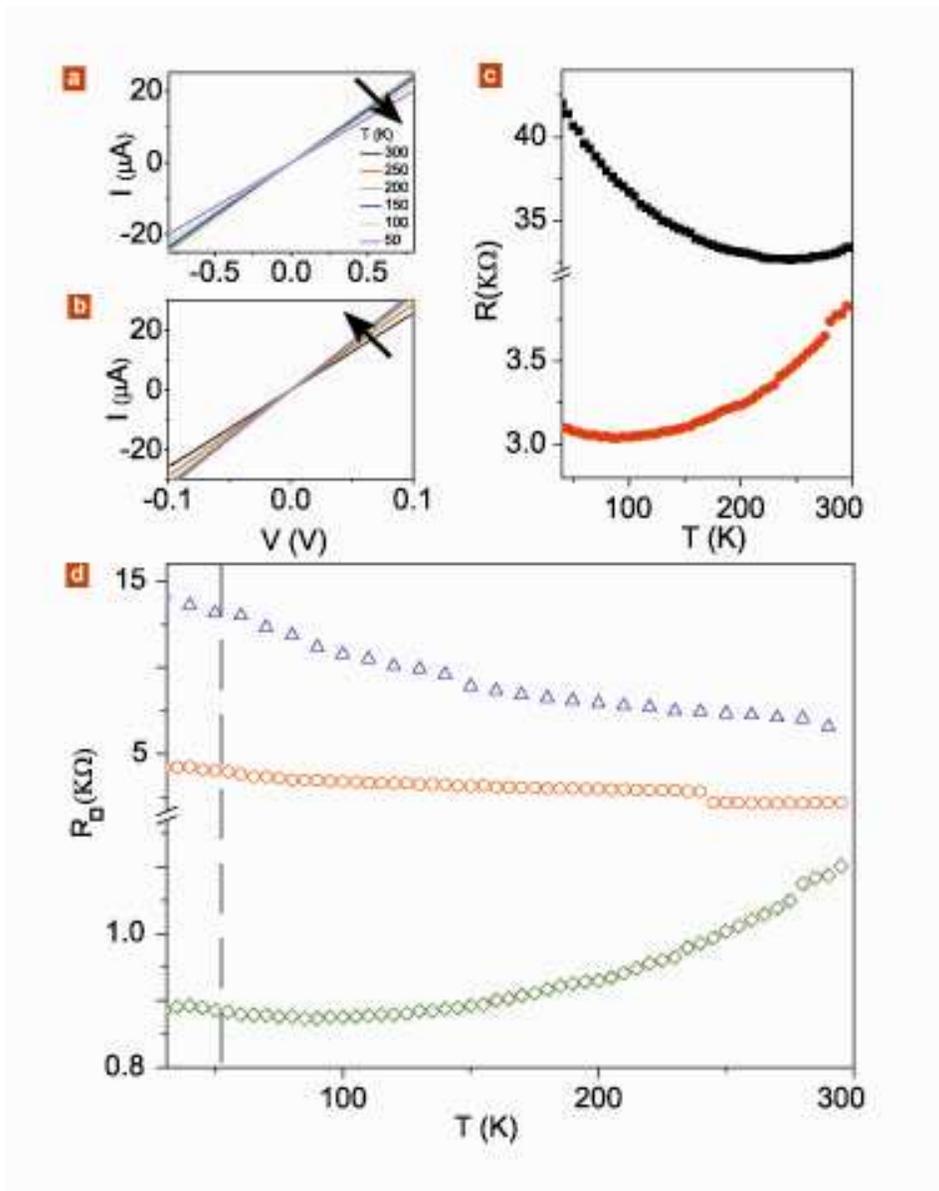

Figure 5 **Temperature dependence of the electrical characteristics of TTF-TCNQ interfaces.** The temperature dependence of the electrical characteristics of TTF-TCNQ interfaces is studied by measuring their I-V curves in the range between 40 and 300 K. Panel a) and b) show the evolution of the temperature dependence of the I-V curves measured in a two terminal (panel a) and four terminal (panel b) configuration. The arrows indicate the variation in current with lowering T. The full temperature dependence of the resistance is shown in panel c). In a two-terminal configuration the resistance is always observed to increase with lowering temperature, contrary to what is observed in four-terminal measurements. Panel d) shows the temperature dependence of the resistance-per-square, measured in a four terminal configuration, in three different samples. The increase in resistance with lowering temperature which is observed in devices with larger square resistance crosses over smoothly into a metallic behavior (decreasing of resistance with lowering temperature) for devices whose square resistance is sufficiently low. The crossover occurs when the resistance-per-square is approximately 4-5 k$\Omega$, of the order of the resistance quantum. The vertical line at T=54 K represents the temperature at which a charge density wave transition is observed in bulk TTF-TCNQ, causing the resistance to increase by a factor of 20 in a range of few degrees Kelvin. In all of our devices, no indication of a resistance increase is visible around this temperature, illustrating the role of the different dimensionality of interfaces as compared to TTF-TCNQ bulk.